\documentclass[twocolumn,british,american]{revtex4-1}
\usepackage[T1]{fontenc}
\usepackage[latin9]{inputenc}
\usepackage{babel}
\usepackage{float}
\usepackage{amsmath}
\usepackage{graphicx}
\usepackage[unicode=true,
 bookmarks=false,
 breaklinks=true,pdfborder={0 0 0},backref=false,colorlinks=false]
 {hyperref}

\makeatletter
\newenvironment{lyxlist}[1]
{\begin{list}{}
{\settowidth{\labelwidth}{#1}
 \setlength{\leftmargin}{\labelwidth}
 \addtolength{\leftmargin}{\labelsep}
 }}
{\end{list}}

\makeatother

\begin{document}

\title{Quantitative analyses of the plant cytoskeleton\\
reveal underlying organizational principles}

\author{David Breuer$^{1}$, Alexander Ivakov$^{2}$, Arun Sampathkumar$^{3}$,
Florian Hollandt$^{1}$, Staffan Persson$^{2,4}$, Zoran Nikoloski$^{1,*}$}

\selectlanguage{british}%

\affiliation{%
\mbox{%
$^{1}$Systems Biology and Mathematical Modelling, Max Planck Institute
of Molecular Plant Physiology,%
}\\
Am Muehlenberg 1, 14476 Potsdam, Germany}

\affiliation{%
\mbox{%
$^{2}$Plant Cell Walls, Max Planck Institute of Molecular Plant Physiology,%
}\\
Am Muehlenberg 1, 14476 Potsdam, Germany}

\selectlanguage{american}%

\affiliation{%
\mbox{%
$^{3}$Sainsbury Laboratory, University of Cambridge,%
}\\
Bateman Street, Cambridge CB2 1LR, United Kingdom}

\selectlanguage{british}%

\affiliation{%
\mbox{%
$^{4}$ARC Centre of Excellence in Plant Cell Walls, School of Botany,
University of Melbourne,%
}\\
Grattan Street, Parkville, Victoria 3010, Australia}

\selectlanguage{american}%

\affiliation{$^{*}$\href{mailto:nikoloski@mpimp-golm.mpg.de}{nikoloski@mpimp-golm.mpg.de}}
\begin{abstract}
The actin and microtubule cytoskeletons are vital structures for cell
growth and development across all species. While individual molecular
mechanisms underpinning actin and microtubule dynamics have been intensively
studied, principles that govern the cytoskeleton organization remain
largely unexplored. Here, we captured biologically relevant characteristics
of the plant cytoskeleton through a network-driven imaging-based approach
allowing to quantitatively assess dynamic features of the cytoskeleton.
By introducing suitable null models, we demonstrate that the plant
cytoskeletal networks exhibit properties required for efficient transport,
namely, short average path lengths and high robustness. We further
show that these advantageous features are maintained during temporal
cytoskeletal re-arrangements. Interestingly, man-made transportation
networks exhibit similar properties, suggesting general laws of network
organization supporting diverse transport processes. The proposed
network-driven analysis can be readily used to identify organizational
principles of cytoskeletons in other organisms.

~

Keywords: cytoskeletal networks, cytoskeletal transport, plant cell
walls, complex networks, organizational principles
\end{abstract}
\maketitle

\section{Introduction}

Complex systems can be represented by networks that capture the underlying
components, as nodes, and their interactions, as links. Network representations
have provided insights into the organizational principles of a variety
of systems, ranging from man-made to systems shaped by evolution,
such as: metabolic networks \cite{Fell2000,Jeong2000}, neural networks
\cite{Felleman1991,Koch1999}, food webs \cite{Williams2000,Jordano2003},
and transportation systems, including: vascular \cite{Gazit1995,West1997}
and leaf venation networks \cite{West1999,Katifori2010}. 

The cytoskeleton represents yet another type of biological network.
It is composed of actin filaments (AFs), microtubules (MTs), and intermediate
filaments that form intricate interconnected arrays. Plant cells lack
intermediate filaments, and their actin and microtubule cytoskeleton
exhibits structural and functional differences to that of animal and
yeast cells. These differences may be due to the presence of a rigid
cell wall, a large central vacuole, the absence of discrete cytoskeleton
organizing centers, or the general need of plants, as sessile organisms,
to cope with changing environmental conditions \cite{Wasteneys2000,Ehrhardt2006}. 

In plant interphase cells, AFs exhibit extraordinarily dynamic behaviors
\cite{Staiger2009}. A major function of the actin cytoskeleton is
to support cytoplasmic streaming, the directed flow of cytosol and
organelles, which is mainly powered by ATP-driven myosin movement
of compartments along the actin cytoskeleton \cite{Shimmen2004,Szymanski2009}.
Furthermore, recent studies have shown that transportation of organelles
depends on the micro-environment of the actin structures, where organelles
are rapidly transported along actin bundles and display reduced motility
when surrounded by thin AFs \cite{Akkerman2011,Sampathkumar2013a}. 

The behavior of single MTs, as well as MT arrays, has been described
throughout the cell cycle and for different cell types \cite{Wasteneys2009,Buschmann2011,Jacques2013a}.
Their dynamics has been well-characterized \cite{Hush1994,Ehrhardt2006,Ehrhardt2008},
and can lead to the formation of self-organized patterns that largely
explain the MTs' orientation in growing cells \cite{Zumdieck2005,Allard2010,Tindemans2010a,Lindeboom2013a}.
While MTs sustain vesicle motility in certain plant cells \cite{Collings2008},
they are typically located at the cell cortex and support the synthesis
of cellulose microfibrils in interphase cells \cite{Paredez2006}.
Nevertheless, there is emerging evidence for transport along MTs also
in these cell types, e.g., Golgi and small cellulose containing compartments
have been reported to track along cortical MTs \cite{Crowell2009}. 

Several studies have investigated the mechanical properties of the
cytoskeleton in yeast and animal cells both experimentally \cite{Wagner2006,Lieleg2007a}
and theoretically \cite{MacKintosh1995,Wagner2006,Benetatos2007}.
Models of AFs as a system of stiff, spring-connected rods have demonstrated
a percolation-related transition in the viscoelastic properties \cite{Ziemann1994,Forgacs1995},
similar to signal propagation in a cytoskeleton model of connected
rods \cite{Shafrir2000,Shafrir2002}. 

The above studies employed a bottom-up approach in which the behavior
of a system is explained based on the dynamics of its components.
However, the interconnected structure and the rapid dynamics of the
cytoskeleton lend themselves to a top-down approach, which is independent
of detailed molecular knowledge and better suited for uncovering the
principles underlying cytoskeletal organization. Two studies have
used such an approach in animal systems: AF arrays have been described
as a superposition of different tessellation models \cite{Fleischer2007}
and a theoretical investigation of cytoskeletal transport in a system
with passive diffusion and active transport along a random network
of segments has demonstrated different regimes of transport \cite{Neri2013}.
Therefore, there is need for a network-based representation of the
cytoskeleton that: (1) captures its complex network structure, (2)
is based on biologically solid ground, (3) can be used to describe
dynamic network processes, and (4) may uncover organizational principles
of the cytoskeleton in plant and animal cells.

In this study, we propose a novel framework that captures the structure
and dynamics of the actin and microtubule cytoskeleton as complex
networks. We used this framework to quantify and compare the behavior
of AFs and MTs in plant interphase cells under different conditions.
We tested the hypothesis that the cytoskeleton is well-suited to support
transport processes. By developing suitable null models as references,
we show that the cytoskeleton indeed exhibits biologically desirable
transport-related properties, such as short average path lengths and
high robustness against disruptions of the network. Finally, we demonstrate
that man-made transportation networks display similar properties.
The developed framework is readily applicable to study the cytoskeleton
of other organisms or under different conditions.

\section{Results}

\begin{figure*}
\begin{centering}
\includegraphics[width=1\textwidth]{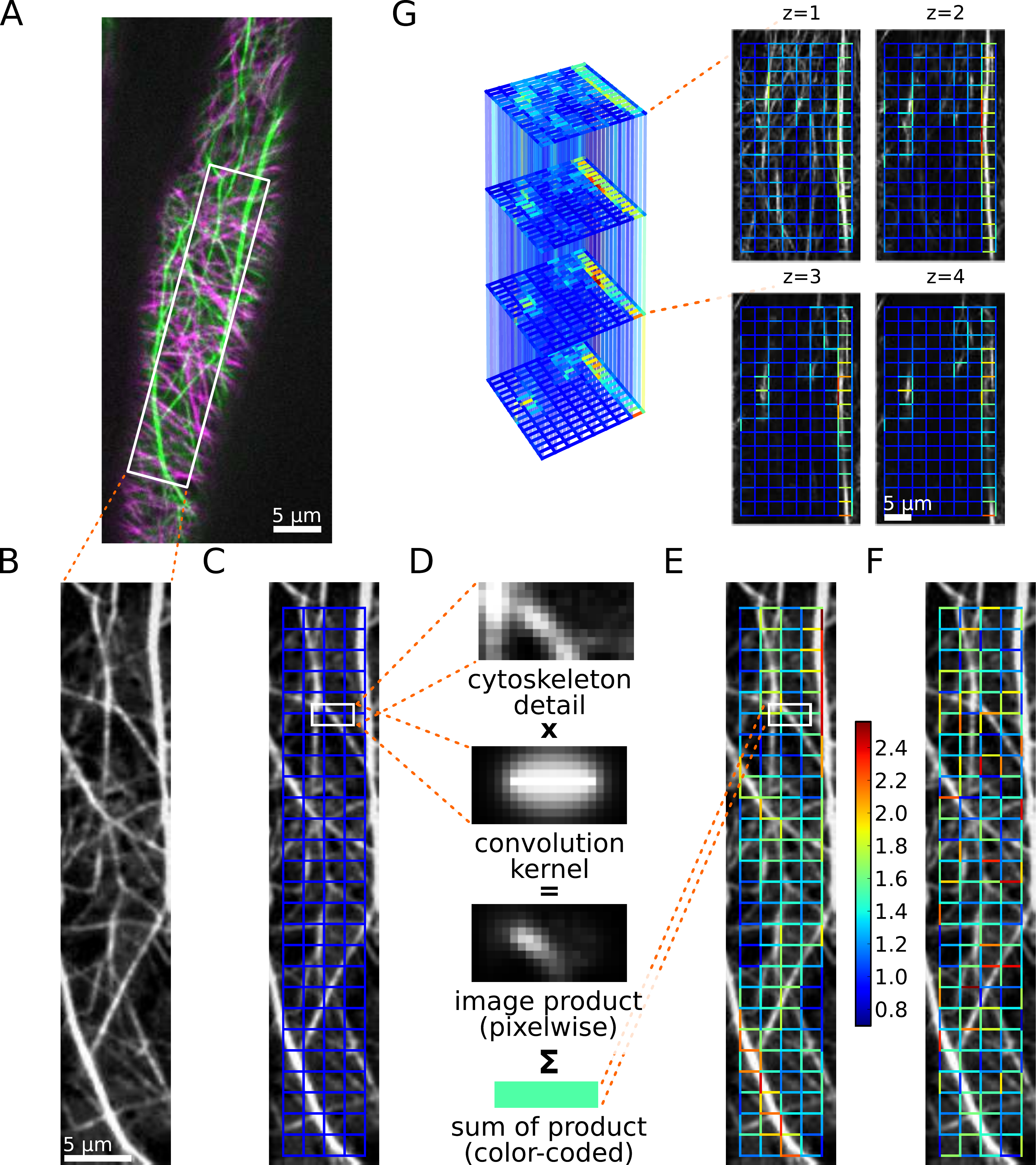}
\par\end{centering}

\protect\caption{\textbf{\label{fig:network}From fluorescence image to reconstructed
network, null model, and 3D extension.} \textbf{(A)} Colored overlay
of unprocessed snapshots of the AFs (green) and MTs (magenta) of a
dual-labeled, three-day-old A.~thaliana hypocotyl cell. \textbf{(B)}
Single preprocessed image of one cytoskeletal component, here AFs.
\textbf{(C)} Grid used for network reconstruction (a uniform rectangular
grid with $10\,\mathrm{pixels}$ spacings is shown in blue). \textbf{(D)}
Using convolution kernels, the links of the grid are assigned scalar
values by pixelwise multiplication of the kernel with the cytoskeleton
image and subsequent summation. \textbf{(E)} Weighted, undirected
network with edges given by the links of the chosen grid type and
weights obtained via the kernel method (weights are color-coded from
blue to red). \textbf{(F)} To assess the biological relevance of various
properties of the cytoskeletal network, a null model is introduced
through an ensemble of networks with shuffled edge weights (one exemplary
realization is depicted). \textbf{(G)} From confocal z-stack recordings,
a three-dimensional cytoskeletal network is reconstructed (grid spacings
are $20\,\mathrm{pixels}$; edges connecting different z-layers are
set transparent for better visibility of the full network).}
\end{figure*}

\subsection{Reconstruction of complex networks from cytoskeletal images}

To investigate the networks of AFs and MTs, we grew Arabidopsis (\emph{A.~thaliana})
FABD:GFP and TUA5:mCherry dual-labeled seedlings \cite{Sampathkumar2011}
in the dark and imaged elongating hypocotyl cells. To capture rapid
changes and to minimize bleaching, we used a spinning-disk confocal
microscope (Fig.~\ref{fig:network}A and B; see electronic supplemental
material (ESM) S1). To generate complex networks from the cytoskeleton
image series we followed a two-step procedure: We placed a grid over
the cytoskeleton which covers the cell's cytoskeleton (e.g.\emph{~}Fig.~\ref{fig:network}C).
From the grid, we constructed an edge-weighted network in which nodes
represent the grid's junctions, and edges represent the grid's links.
We assigned a weight to each edge by creating convolution kernels
with Gaussian profiles for each edge (Fig.~\ref{fig:network}D),
thus projecting the cytoskeleton onto the overlaid grid. This results
in a weighted, undirected network (Fig.~\ref{fig:network}E) where
the weights reflect the intensity of the underlying filaments/bundles.
Using confocal z-stack image series, these steps were also used to
construct three-dimensional cytoskeletal networks (Fig.~\ref{fig:network}G).
The procedure was repeated for all images of the recorded actin and
microtubule time series, separately. As a result, each network captures
information of the time-dependent cytoskeletal component whose properties
may be readily investigated. 

To determine if the studied network properties carry a biological
signal, we developed several null models that randomize parts of the
cytoskeletal structures while preserving the total amount of cytoskeleton
in the cell (cf\emph{.~}Appendix \ref{sec:app_2}). If a given network
property is significantly higher or lower than expected by chance
we conclude that the underlying cytoskeletal organization is non-random
and, therefore, biological relevant. This may suggest that the cytoskeleton
is tuned to guarantee such values of the structural or functional
network property.

\begin{figure*}
\begin{centering}
\includegraphics[width=1\textwidth]{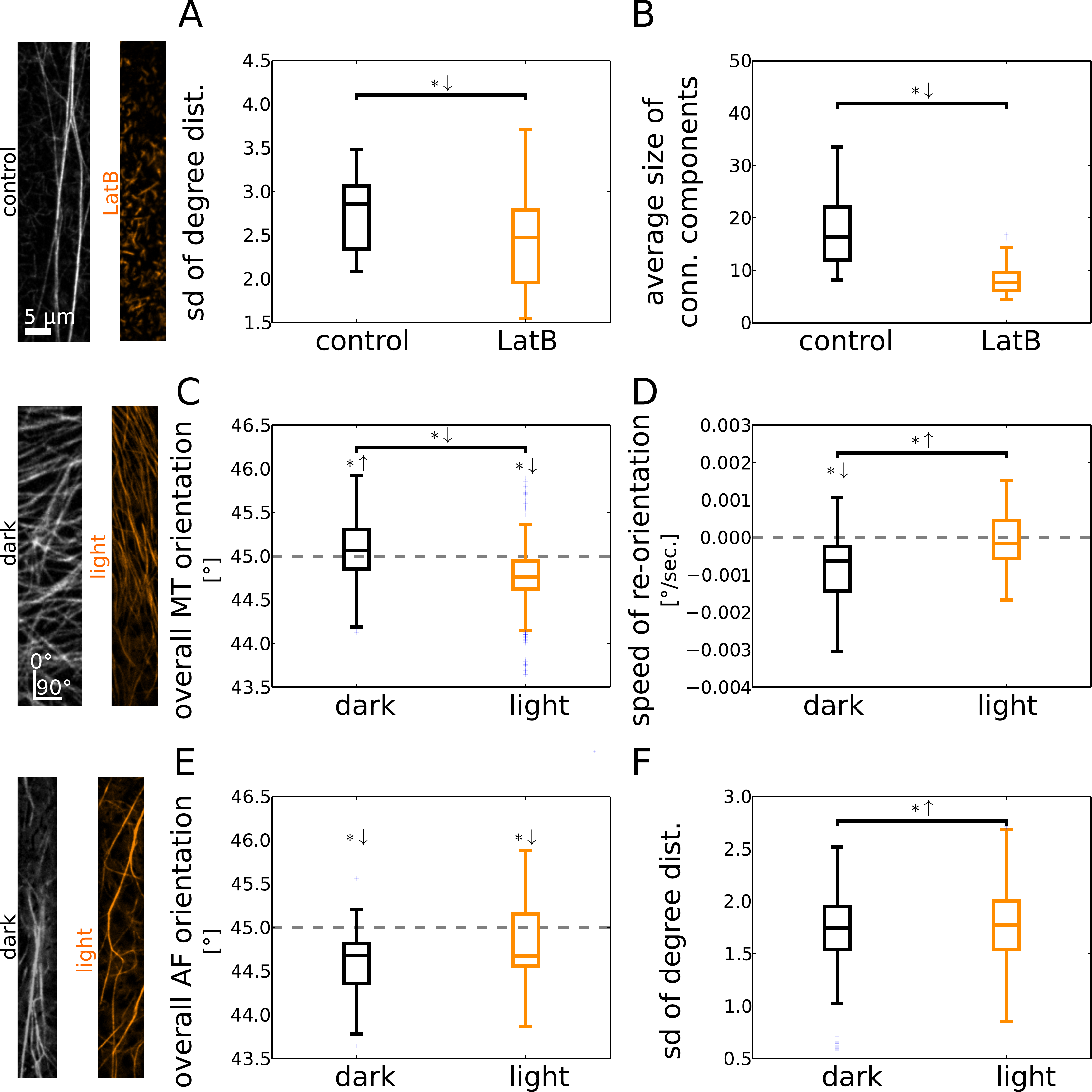}
\par\end{centering}

\protect\caption{\textbf{\label{fig:quantification}Network properties capture biologically
relevant aspects of cytoskeletal organization for different scenarios.}
The first 20 frames of the image series are used for the analysis.
$*\downarrow$ and $*\uparrow$ above a bar denote a decrease or increase
of the properties of the treated relative to those of the control
plants (independent two-sample $t$-test: $p\mathrm{-value}<0.05$).
$*\downarrow$ and $*\uparrow$ above a box plot denote network properties
that fall below or exceed a reference value marked by a gray, dotted
line (one-sample two-sided $t$-test: $p\mathrm{-value}<0.05$). \textbf{(A)}
The AF network of Latrunculin B-treated plants displays a smaller
standard deviation of the network\textquoteright s degree distribution.
\textbf{(B)} The connected patches of AFs are smaller after Latrunculin
B treatment. \textbf{(C)} The orientation of MTs is predominantly
horizontal in dark-grown plants and vertical in light-exposed plants.
\textbf{(D)} Computing the change in MT orientation per unit time
shows a difference between control and light-treated plants. In dark-grown
plants, a significant change towards a vertical orientation is observed,
which is absent for light-treated plants. \textbf{(E)} The (horizontal)
orientation of AFs is not altered in plants exposed to light. \textbf{(F)}
Light induces a dispersion of AFs which yields a broader degree distribution.}
\end{figure*}

\subsection{The reconstructed networks capture biologically relevant features
of the actin and microtubule cytoskeletal components}

To test whether the proposed network-based approach captures biologically
meaningful features, we used chemical treatments and environmental
stimuli to alter the behavior of the cytoskeleton. First, we quantified
the effect of the actin-disrupting drug Latrunculin B on the actin
cytoskeleton. This drug binds to monomeric actin and, thereby, inhibits
AF formation \cite{Yarmola2000}. We reconstructed the AF networks
for both the control and the treated plants for each frame of the
image series. The structure of the AFs and their drug-induced fragmentation
was quantified by two network properties which can be related to the
biological phenomenon (see Appendix \ref{sec:app_3} for a mathematical
description and detailed interpretation of these quantities): The
standard deviation of the degree distribution \cite{West2001} captures
the spatial heterogeneity of the distribution of actin structures,
i.e., images with regions of low and high cytoskeletal intensities
yield both small and large edge weights and consequently a broader
degree distribution (cf\emph{.~}Fig.~\ref{fig:network}E). Since
the edge weights integrate intensities of possibly multiple filaments,
our approach does not resolve differences in thicknesses or numbers
of individual filament but only a combination thereof. By comparing
the standard deviations of the degree distributions of control and
treated plants, we found a statistically significant reduction by
Latrunculin B (Fig.~\ref{fig:quantification}A; independent two-sample
$t$-test: $p\mathrm{-value}=7.0\cdot10^{-9}$; for treated and non-treated
plants, respectively, we pooled the standard deviations of the degree
distributions across the first 20 time points of the image series).
We then determined the average number of nodes per connected, non-trivial
network component after thresholding the edge weights, providing an
estimate for the extent to which the cytoskeletal filaments form connected
networks. By using the $50$th percentile as a threshold, we found
that Latrunculin B reduces the average size of the resulting connected
components (Fig.~\ref{fig:quantification}B; independent two-sample
$t$-test: $p\mathrm{-value}=2.9\cdot10^{-42}$). These findings are
in agreement with visual reports on the fragmented actin structure
of Latrunculin B-treated cytoskeletons \cite{Yarmola2000}.

By using the reconstructed MT network, we quantified the overall orientation
of MTs in plants that had been exposed to light several hours before
imaging. Light is one of the environmental factors that determine
plant growth, and it is well established that the MT array rapidly
changes from largely transverse to a generally longitudinal when seedlings
are exposed to light \cite{Wymer1996,Paradez2006,Sambade2012}. As
our method does not detect individual filaments, we inferred the MT
orientation indirectly (see Appendix \ref{sec:app_4} for a detailed
derivation): By placing an imaginary rod of a specific length and
orientation over the grid we calculated its contributions to the weights
of edges with different orientations by using our kernel method (cf\emph{.~}Fig.~\ref{fig:network}D).
Here, we solved the inverse problem to obtain the overall MT orientation
$\alpha$ from the weight distribution of edges with different orientations
. Angles $\alpha\in\left[0^{\circ},45^{\circ}\right)$ and $\alpha\in\left(45^{\circ},90^{\circ}\right]$
indicate overall vertical and horizontal orientations of the MTs,
respectively. We estimated the MT orientation for seedlings grown
under dark and light conditions and found a significant difference
(Fig.~\ref{fig:quantification}C; independent two-sample $t$-test:
$p\mathrm{-value}=5.8\cdot10^{-52}$) with a horizontal and longitudinal
orientation, respectively (one-sample two-sided $t$-tests: dark $p\mathrm{-value}=6.3\cdot10^{-8}$,
light $p\mathrm{-value}=4.7\cdot10^{-45}$). These findings are in
agreement with known results \cite{Granger2001,Paradez2006,Sampathkumar2011,Sambade2012}.
They further revealed that despite the strong correlation between
light exposure and longitudinal MT orientation, there are also deviants
(cf\emph{.~}Fig.~\ref{fig:quantification}C): Under dark condition,
a fraction of about 40\% of the MT networks shows an, unexpected,
overall vertical orientation while under light conditions, about 20\%
of the analyzed MT networks display an overall horizontal orientation,
contrary to expectations. . These deviations from the expected results
may highlight the inherent variability of cytoskeletal responses to
external stimuli and support the view that hypocotyl cells may be
in different stages of growth \cite{Gendreau1997,Le2005,Sambade2012}.

We also studied the speed of MT reorientation under the microscope
by computing the slope of the average orientation time series via
a linear regression. There was a significant difference between the
two treatments (Fig.~\ref{fig:quantification}D; independent two-sample
$t$-test: $p\mathrm{-value}=2.5\cdot10^{-3}$), i.e., while the change
in orientation is negative in dark, it does not significantly differ
from zero in light (one-sample two-sided $t$-test: dark $p\mathrm{-value}=2.5\cdot10^{-5}$,
light $p\mathrm{-value}=0.6$). Therefore, we conclude that over the
range of five minutes of the experiment the confocal laser light does
induce a reorientation of the MTs towards the longitudinal cell axis
in dark-grown plants; however, this is not the case in plants exposed
to light before imaging as reorientation of MTs has already progressed
further in these cells. 

Next, we employed our network-based framework to investigate the behavior
of AFs in response to light in growing hypocotyls. Like for MTs, we
inferred the overall orientation of the AFs for dark-grown and light-treated
plants, respectively. We found no significant difference in actin
orientation between the treatments (Fig.~\ref{fig:quantification}E;
independent two-sample $t$-test: $p\mathrm{-value}=6.6\cdot10^{-2}$),
with a consistent major longitudinal orientation (one-sample two-sided
$t$-test: dark $p\mathrm{-value}=3.8\cdot10^{-7}$, light $p\mathrm{-value}=4.4\cdot10^{-2}$).
To quantify the heterogeneity of the actin distribution, we computed
the standard deviation of the degree distributions for both scenarios.
In light-treated plants, the actin cytoskeleton displayed a more heterogeneous
distribution across the cell than in dark-grown plants (Fig.~\ref{fig:quantification}F;
independent two-sample $t$-test: $p\mathrm{-value}=1.5\cdot10^{-4}$),
implying the prevalence of bundles. These findings agree with reports
on the impact of light on the organization of AFs in maize coleoptiles
\cite{Waller1997}. However, they do not agree with the qualitative
findings in a different species, i.e., rice, where light was shown
to promote a change in AF orientation from transverse to longitudinal
and to disperse actin bundles \cite{Holweg2004}. Interestingly, the
rearrangement of AFs under light has been linked to that of MTs \cite{Sampathkumar2011}.
Therefore, our findings suggest that a change in environmental conditions
would impose a need for rapid redistribution of cellular material
in the cell, which are known to be facilitated by actin bundles \cite{Akkerman2011}. 

\begin{figure*}
\begin{centering}
\includegraphics[width=1\textwidth]{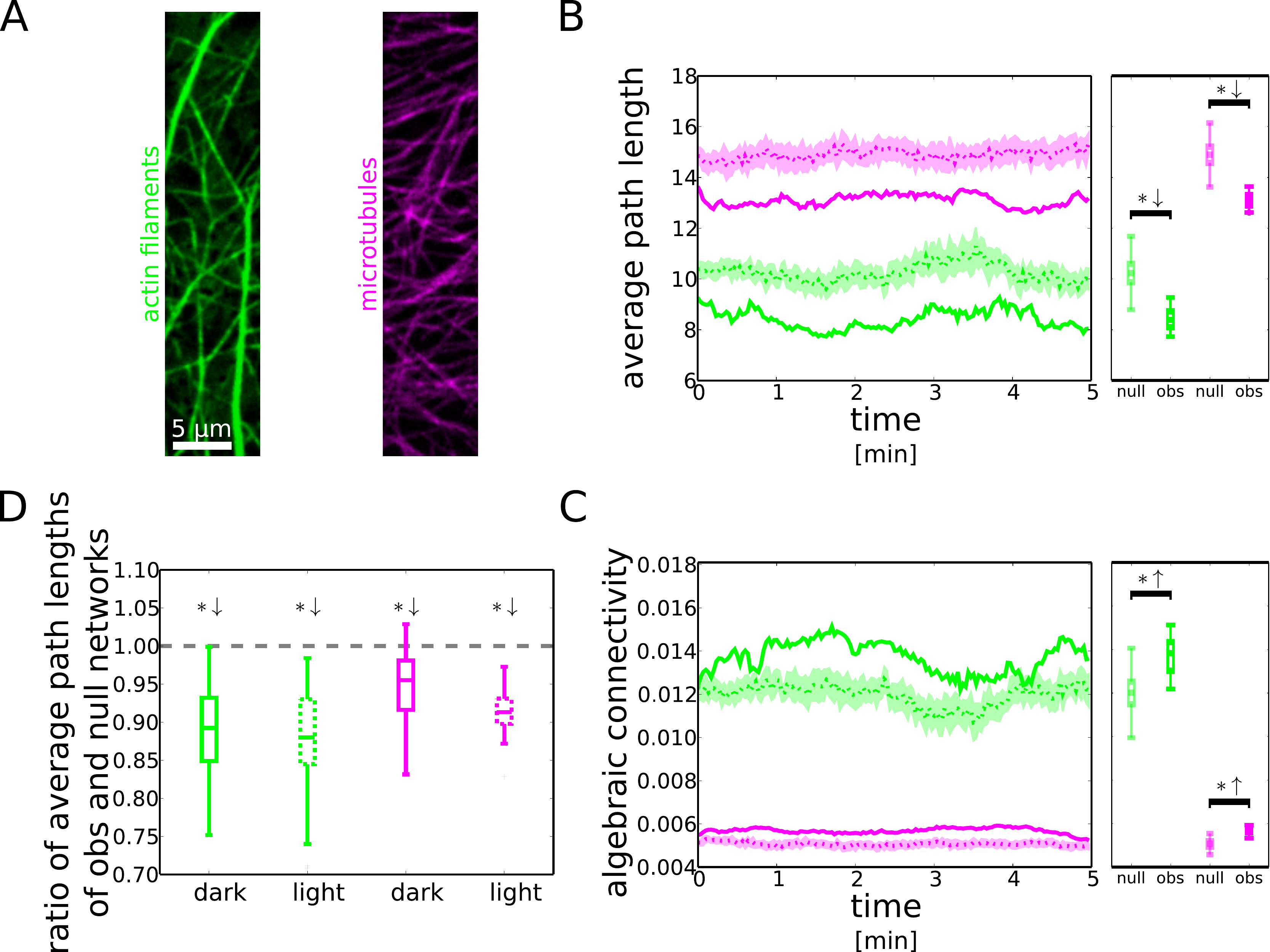}
\par\end{centering}

\protect\caption{\textbf{\label{fig:efficiency}Time-resolved average path length and
algebraic connectivity of cytoskeletal networks and null model networks.}
In (B) and (C), the results for the observed networks (solid lines)
of AFs (green) and MTs (magenta) are compared to those of the null
model (dashed lines: ensemble-mean; shaded regions: ensemble-mean
$\pm$ standard deviation). The box plots show the distribution of
values of a network property used in the statistical test. \textbf{(A)}
Green- and magenta-colored images of AFs and MTs, respectively. \textbf{(B)}
The average path lengths of AF and MT networks (solid lines) fluctuate
over time and stay well below the average path lengths of the null
model (dashed lines and shaded regions). \textbf{(C)} The algebraic
connectivity is consistently larger for both AFs and MTs in the observed
cytoskeletal networks (solid lines) than in the null model networks
(dashed lines and shaded regions). \textbf{(D)} Comparing the ratios
of average path lengths of the observed networks and their null model
networks for AFs and MTs yields no statistically significant difference
between dark and light conditions.}
\end{figure*}

\subsection{Accessibility and robustness of cytoskeletal networks}

After demonstrating the inherent ability of our network approach to
capture biologically relevant information on cytoskeletal organization,
we focused on identifying network properties that reflect the functions
of the cytoskeleton. To investigate the transport efficiency of the
AF and MT networks (Fig.~\ref{fig:efficiency}A), we computed average
path length (as a measure for the cellular accessibility of the cytoskeleton)
and algebraic connectivity (as a measure for the cytoskeletons' robustness
against disruptions) (see Appendix \ref{sec:app_3} for the mathematical
formulation and a detailed interpretation of the properties). 

The average (shortest) path length (APL) \cite{West2001} is the average
of the minimum distances between all pairs of nodes in a (edge-weighted)
network. Here, the length of an edge is given by the inverse of its
weight, i.e., thick actin bundles or tubulin filaments yield small
edge lengths. This is reasonable since cytoskeletal bundles typically
facilitate faster transport compared to thinner filaments \cite{Akkerman2011}
which may, in general, depend on the size of the cargo. The APL provides
an estimate of how close any two nodes are expected to be and, hence,
the accessibility in the cytoskeleton. By computing the APLs for the
sequence of the AF and MT networks, we obtained two time series (Fig.~\ref{fig:efficiency}B,
green and magenta, solid lines). The corresponding values largely
reflect the overall intensity distribution of the images and, by themselves,
carry little information about the underlying network structure. As
a reference, we calculated the APLs for ensembles of AF or MT null
model networks, i.e., networks obtained by shuffling edge weights
(Fig.~\ref{fig:efficiency}B, green and magenta, dashed lines and
shaded regions). We found that the observed networks exhibit significantly
smaller APLs than their respective null models (independent two-sample
$t$-test: AF $p\mathrm{-value}=7.8\cdot10^{-273}$, MT $p\mathrm{-value}<2.2\cdot10^{-308}$). 

The algebraic connectivity (AC) \cite{West2001} is the second smallest
eigenvalue of the network's graph Laplacian, which is closely related
to the weight matrix of the network, and reflects how well-knit the
network is. While a vanishing AC indicates the decomposition of the
network into two or more disconnected components, larger values correspond
to a higher robustness of the network against disruptions. By comparing
the AC of the observed actin and microtubule networks (Fig.~\ref{fig:efficiency}C,
green and magenta, solid lines) to their null model counterparts (Fig.~\ref{fig:efficiency}C,
green and red, dashed lines and shaded regions), we found that the
observed networks yield significantly larger algebraic connectivities
than their respective null model networks (independent two-sample
$t$-test: AF $p\mathrm{-value}=7.5\cdot10^{-139}$, MT $p\mathrm{-value}=2.9\cdot10^{-278}$). 

These findings may be graphically explained as follows: Networks of
MTs, and even more so AFs, possess filaments and bundles that stretch
across large parts of the cell. These structures establish connected
paths in the networks with large weights and therefore small lengths.
In the computation of shortest path lengths, they act as ``highways\textquotedblright{}
that efficiently connect spatially distant regions. Furthermore, the
AF and MT networks exhibit larger regions that are particularly strongly
linked and result in a higher robustness of the networks against disruption.
To support this interpretation, we computed the degree assortativity
\cite{Newman2009a} given by the correlation between degrees of nodes
and those of their neighbors. It quantifies the extent to which nodes
of (dis-)similar degree are connected to each other. Both the AF and
MT networks exhibit significantly higher assortativity than their
corresponding null model networks and are hence more spatially clustered
(independent two-sample $t$-tests: AF $p\mathrm{-value}<2.2\cdot10^{-308}$,
MT $p\mathrm{-value}<2.2\cdot10^{-308}$).We note that both APL and
AC are summary statistics that do not capture differences in local
connectivity patterns but reflect network properties that relate to
the network's overall transport capacity.

Interestingly, despite the differences in the network architecture
of AFs and MTs under dark and light conditions (cf\emph{.~}Fig.~\ref{fig:quantification}),
there were no significant differences in the ratios of APLs of observed
and null model networks (Fig.~\ref{fig:efficiency}D). Moreover,
these ratios stay consistently below one throughout the time series
(one-sample two-sided $t$-tests: $p\mathrm{-values}<0.05$ for AF/MT
dark/light), reflecting small effort for reaching any part of the
cytoskeletal networks. Thus, the cytoskeleton preserves its advantageous
transport properties over time and across conditions. 

While the MT network can largely be captured at the cell cortex in
interphase cells, the actin cytoskeleton constitutes a three-dimensional
structure that spans the expanding cell. To ensure that we captured
the volumetric behavior of the AFs, we also recorded confocal z-stack
image series of such cells and used our framework to reconstruct the
AF network as a three-dimensional network (cf\emph{.~}Fig.~\ref{fig:network}G).
To assess if the additional information about AFs below the cortical
plane changes the transport efficiency of the AF network, we compared
the APL and AC of the three-dimensional network to that of the two-dimensional
network obtained by averaging the intensities of edges at the same
x-y-position across all z-layers. In both cases, the ratio of network
properties of the observed network and the null model networks stays
well below one (one-sample two-sided $t$-test: $p\mathrm{-values}<0.05$
for 2D/3D APL/AC; see Appendix \ref{sec:app_2} for details). Analogously
to the bundle structures in the 2D networks, the actin structures
that reach deeper into the cell provide strong connections in the
3D reconstruction (see Fig.~\ref{fig:network}G) that naturally equip
the network with shorter paths and higher robustness. 

Taken together, AF as well as MT networks display highly non-random
features. In particular, short APL (good accessibility) and large
AC (high robustness) are preserved over time and across different
environmental conditions. These findings provide quantitative support
for the idea that plants reliably establish and maintain cytoskeletal
structures that are optimized for transport processes throughout the
cell \cite{Shimmen2004,Paredez2006,Sampathkumar2013a}.

\begin{figure*}
\begin{centering}
\includegraphics[width=1\textwidth]{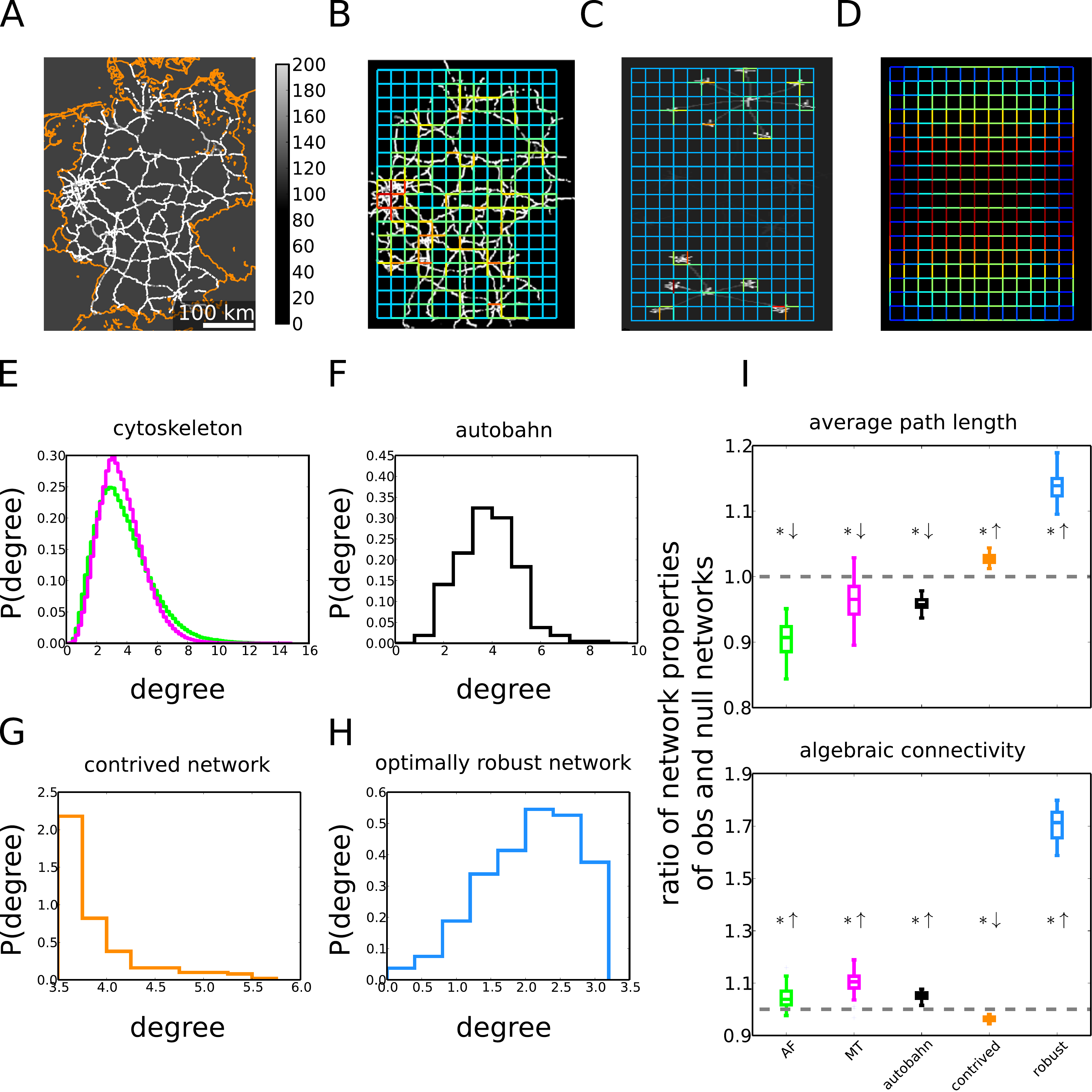}
\par\end{centering}

\protect\caption{\textbf{\label{fig:comparison}Comparison of cytoskeleton, German
autobahn, and other types of networks.} \textbf{(A)} The autobahn
network of Germany with color-coded speed limits and national borders
(orange) for guidance. \textbf{(B)} Network reconstructed from the
autobahn image. \textbf{(C)} Reconstructed network for a contrived
underlying structure with different structural and transport-related
properties. \textbf{(D)} Network with maximal algebraic connectivity
for the given grid-topology and normalized sum of weights (here, the
network is constructed via an optimization procedure and not inferred
from an underlying image intensity distribution; the black background
is for better visibility only). \textbf{(E)} The degree distributions
of the AF (green) and MT (magenta) networks are unimodal and peak
around their means (excess kurtosis $>0$; the resolution of the histograms
is higher as they include all networks of an image series, cf\emph{.}~Fig.~\ref{fig:efficiency}).
\textbf{(F)} The degree distribution of the autobahn network is unimodal
and peaks around its mean (excess kurtosis $>0$). \textbf{(G)} For
the contrived network, the degree distribution is also peaked (excess
kurtosis $>0$) but shows a more heavy right tail compared to (E)
and (F). \textbf{(H)} The degree distribution of the optimally robust
network is broader around the peak and has thinner tails (kurtosis
$<0$). \textbf{(I)} The ratios of average path lengths of observed
and null model networks are below one for the cytoskeletal and the
autobahn networks and above one for the contrived and the optimally
robust network. The algebraic connectivities are bigger than expected
by chance for all studied networks, except for the contrived networks
that show a smaller algebraic connectivity. All deviations from the
unit ratio are statistically significant (one-sample two-sided $t$-tests:
all $p\mathrm{-values}<0.05$). }
\end{figure*}

\subsection{The cytoskeleton and the German autobahn exhibit similar network
properties}

We next asked whether the observed efficiency in network properties
is unique to the cytoskeleton or if it can also be observed in other
transportation networks. As a prominent example, we generated images
of the German autobahn with color-coded speed limits (Fig.~\ref{fig:comparison}A;
The maximum speed limit is set to $200\,\mathrm{km/h}$, the speed
limit outside of the autobahn is set to $50\,\mathrm{km/h}$. However,
our findings are largely independent of this choice, see Appendix
\ref{sec:app_5}). By using our established framework, we obtained
a weighted network for the autobahn (Fig.~\ref{fig:comparison}B)
and corresponding null model networks by shuffling edge weights. For
the APL and the AC, we computed the ratios for the respective network
properties of the autobahn and its null model networks and we found
that the autobahn network exhibits shorter APLs and higher algebraic
connectivities than expected by chance (Fig.~\ref{fig:comparison}I,
black). These results are similar to what we found for the cytoskeleton
(Fig.~\ref{fig:comparison}I, green and magenta). Furthermore, we
note that the degree distributions of the cytoskeletal networks and
the autobahn are unimodal and peak around their means (Fig.~\ref{fig:comparison}E
and F). 

To differentiate these networks from networks with different structural
and transport-related properties, we further studied a contrived network
with a stronger local structure and weaker long-ranged connections
(Fig.~\ref{fig:comparison}C). The contrived network displays a heavy-tailed
degree distribution (Fig.~\ref{fig:comparison}G) as well as properties
associated with poor transport efficiency, namely, longer APL and
smaller AC than expected from the null model (Fig.~\ref{fig:comparison}I,
orange). 

Another interesting comparison is that of the cytoskeleton and the
autobahn to networks in which one or several transport-related properties
are optimized. The weight distribution of a network with a fixed sum
of edge weights and maximal AC may be computed efficiently by solving
a semi-definite optimization problem whose solution is unique \cite{boyd2006convex,Sun2006}
(Fig.~\ref{fig:comparison}D). Such an optimally robust network outperforms
the cytoskeletal and the autobahn networks by a factor of eight with
respect to the ratio of AC of observed and null model networks (Fig.~\ref{fig:comparison}I,
blue). However, it is less efficient in terms of its APL, which is
higher than expected by chance, demonstrating a trade-off between
different measures of network optimality.

While there are certainly differences in the structure as well as
the function of the cytoskeleton and the autobahn, the network properties
studied here are summary statistics and it is not possible to infer
local structural differences from them. In particular, we do not study
their absolute values but the relative efficiency of these networks
with respect to their respective null model which may point to its
organizational principles.

To conclude, both the actin and microtubule cytoskeleton display characteristics
typical of transportation networks, such as the autobahn, and exhibit
structures which may not be aimed at optimizing a single property
indicative of efficient transport. Our data therefore provide quantitative
measures to support a view of the plant interphase cytoskeleton as
an efficient transportation network.

\section{Discussion}

Though many studies have analyzed the cytoskeleton, most of them have
relied on qualitative observations or manual tracking of up to some
dozens of AFs and MTs. The rapid dynamics, as well as strong variability
of cytoskeletal organization across different cell types, stages of
cell life, and environmental conditions, necessitate a framework that
allows for a fast and objective quantification of the cytoskeletal
components in living cells. This would then allow for biologically
meaningful interpretations that go beyond strictly theoretical studies
to investigate the structure of the system. Here, we described one
such framework that captures biologically relevant variations.

Many studies have used a bottom-up approach to the cytoskeleton, in
which the molecular principles are presupposed and used to infer the
behavior of the system. Instead, we pursued a top-down strategy to
represent the cytoskeletal organization without the need for detailed
molecular knowledge. Hence, our approach hints at the underlying organizational
principles of the cytoskeleton. More specifically, by choosing a representation
through complex networks, we could exploit the well-equipped toolkit
from graph theory to investigate the structure of the cytoskeleton
and its relation to, e.g., efficient transport processes in the cell.

For a careful interpretation of these findings, we need to bear in
mind several points: (1) Our network reconstruction method creates
nodes at positions given by the chosen grid (cf\emph{.}~Fig.~\ref{fig:network}).
Hence, not all nodes correspond to crossings of the filaments. Moreover,
despite our focus on the largely planar cortical cytoskeletal even
apparently crossing filament maybe separated by hundreds of nanometers
in z-direction. Using our current imaging techniques, such distances
may not be resolved in the 3D reconstruction either (see outlook below).
In particular, such distances prohibit the switching of motor proteins
(cf\emph{.}~e.g\emph{.}~\cite{Balint2013} for MT). Yet, the edge
weights in our method agglomerate local intensities that may originate
from multiple filaments in different depths. More importantly, typical
cargo such as mitochondria, Golgi bodies, or chloroplasts range from
several $100\,\mbox{\ensuremath{\mathrm{nm}}}$ to several $\mathrm{\mu m}$
in size and may thus easily bridge even larger distances between filaments,
thereby justifying our assumption of transport along edges and via
nodes. (2) In our approach, all edges are undirected, i.e., they allow
bi-directional transport. While bi-directional transport may occur
along single actin or microtubular filaments, e.g., due to different
motor proteins or fluctuations \cite{Gross2004,Lee2004,Caviston2006},
bundles of filaments typically allow uni-directional only. This uni-directionality
is further amplified by the cytoplasmic bulk flow generated by the
coordinated movement of motor proteins \cite{Shimmen2004}. Yet again,
our reconstruction methods assigns edge weights by integrating local
intensities of possibly multiple filaments with different orientations.
Thus, fully bi-directional of transport is unlikely (but not excluded)
and, since it can not be inferred from the cytoskeletal images alone,
we use it as an approximation of the potential transport capacity.
(3) This potential transport capacity is modeled to be higher in regions
with many/thick cytoskeletal filaments, as described by the edge weights.
However, the edge weights do not quantify the speed/amount of cargo
that is really transported (and we do not measure it, see outlook
below). (4) Finally, we note that AF and MT networks generally transport
different cargo (cf\emph{.~}\cite{Shimmen2004,Paredez2006}), although
there is evidence for transport of, e.g.\emph{,} small cellulose containing
vesicle along both structures \cite{Goode2000,Gutierrez2009}. In
addition, different types of cells may require different modes of
transportation \cite{Hussey2006}. Here, we focused on the potential
transport capacity of the cytoskeleton in interphase hypocotyl cells,
but our framework may readily be used to study other scenarios. 

There is a rich literature on the comparison of structures of different
networks. Many biological and man-made networks show scale-free degree
distributions, i.e., there are a few nodes with many neighbors \cite{Bornholdt2003},
e.g., airway networks \cite{Guimera2005a}. However, nodes in other
transportation networks are restricted regarding the number of potential
neighbors due to the physical limitations. Road and railway networks
display degree distributions that peak around their average values
\cite{Barthelemy2011}, which we also demonstrated for the cytoskeletal
networks. Despite the apparently diverse principles underlying man-made
transportation networks, studies have revealed strong agreement in
a number of their properties, e.g., degree distribution. This agreement
may be explained by costs associated with the establishment of new
nodes and links \cite{Barthelemy2008,Courtat2011,Louf2013}. Our findings
suggest that comparable cost-related restrictions may play a role
in the formation of the cytoskeleton, leading to similar structures
and transport properties as in man-made networks.

In summary, our framework captures the complex network structure of
filamentous cytoskeletal components. We used this framework to derive
organizational principles of the cytoskeleton. We further showed that
AF and MT networks display biologically desirable characteristics,
such as short APLs and high robustness, similar to characteristics
found in non-biological transportation networks. In particular, these
features of efficient transportation networks are maintained over
time and across conditions.

Possible directions of future efforts are manifold: (1) Our framework
can be employed to quantify the complex structures of AF and MT networks,
and thus enables an automated and objective comparison of the complex
structures of cytoskeletal networks in other biological systems, e.g.,
focusing on the cytoskeleton connecting the nucleus to other parts
of the cell. (2) The resolution of the fine cytoskeletal structures
may be improved by using more advanced imaging techniques like total
internal reflection fluorescence microscopy, at least for the cortical
cytoskeleton. (3) Another promising direction is the comparison of
reconstructed cytoskeletal networks to networks that optimize one
or several seminal network properties. As different network structures
favor specific properties, the cytoskeleton may represent an evolutionarily
shaped compromise between them. While such a balance has been suggested,
e.g., between the speed and the sensitivity in the polarization of
the cytoskeleton \cite{Hawkins2010}, quantitative evidence for a
trade-off in the cytoskeleton\textquoteright s transport properties
is lacking. We note that besides its vital role in cellular transport
processes the plant cytoskeleton strikingly determines the mechanical
properties of the cell. (4) Finally, our work paves the way for direct
studies of the cytoskeleton as a transportation network. Employing
actin and organelle dual-labeled plants, it is appealing to correlate
actual biological transport processes with flow-related network measures.
While several studies have investigated the transport of organelles
and vesicles along the cytoskeleton \cite{Goode2000,Rogers2000,Shimmen2004,Balint2013},
none have quantitatively linked it to the complex structure of the
cytoskeletal network. Answering these questions may contribute to
a better understanding of the organizing and dynamic principles of
the cytoskeleton.

\section{Materials and Methods}

The experiment setup includes dual-labeled \emph{Arabidopsis thaliana}
Columbia-0 seedlings to which different treatments were applied prior
to imaging in a spinning disk confocal microscope setup. For further
details, refer to Appendix \ref{sec:app_1}. The computational network-based
investigation of the image series, as illustrated in Fig.~1, includes:
(1) the preprocessing of the images in Fiji \cite{Schindelin2012},
(2) the creation and quantification of the weighted cytoskeletal and
null model networks, and (3) their statistical analyses in Python
\cite{VanRossum2013} (using SciPy \cite{Jones2001}, NumPy \cite{Oliphant2006},
NetworkX \cite{Hagberg2008} and the Matplotlib \cite{Hunter2007}
libraries). The construction of an optimally robust network was performed
by solving a semi-definite optimization problem using the Cvxopt Python
package \cite{Dahl2006}. Detailed descriptions of these steps are
given in Appendix \ref{sec:app_2} and the studied network properties
are described in detail in Appendix \ref{sec:app_3}. The overall
orientation of cytoskeletal components is inferred from the network's
weight distribution as described in Appendix \ref{sec:app_4}. The
data of the German autobahn, as depicted in Fig.~\ref{fig:comparison},
are collected from OpenStreetMap and filtered as explained in Appendix
\ref{sec:app_5}.

\section{Acknowledgments}

D.B., A.I., S.P., and Z.N.~were supported by the Max-Planck-Gesellschaft.
We thank Dr.~Tijs Ketelaar and Dr.~Georg Basler for valuable comments
on the manuscript.

\section{Abbreviations List}

AF: actin filament, MT: microtubule, APL: average path length, AC:
algebraic connectivity

\bibliographystyle{apalike}

\cleardoublepage
\appendix
\onecolumngrid
\setcounter{page}{1}
\renewcommand{\thepage}{A\arabic{page}}
\setcounter{table}{0}
\renewcommand{\thetable}{A\arabic{table}}
\setcounter{figure}{0}
\renewcommand{\thefigure}{A\arabic{figure}}
\setcounter{section}{0}
\renewcommand{\thesection}{A\arabic{section}}

\section{\label{sec:app_1}Experimental setup}

Here, we describe the experimental setup for recording the cytoskeleton
of growing plant cells. Dual-labeled \emph{Arabidopsis thaliana} Columbia-0
seedlings were were previously described in \cite{Sampathkumar2011}.
The seedlings were surface sterilized (ethanol), stratified for $2\,\mathrm{days}$
at $4^{\circ}\mathrm{C}$ and germinated on MS agar plates (1X Murashige
and Skoog salts, $8\,\mathrm{g/L}$ agar, 1X B5 vitamins, and $10.8\,\mathrm{g}$
$8\,\mathrm{g/L}$ sugar). All plants were grown in the dark on vertical
plates at $21.8^{\circ}\mathrm{C}$ for $3\,\mathrm{days}$. For treatment
with Latrunculin B, seedlings were floated on distilled water containing
$150\,\mu\mathrm{M}$ Latrunculin B and a set of control seedlings
on pure water in $6$-well plates. The seedlings were incubated in
the dark with gentle shaking for $4\,\mathrm{hours}$ before imaging.
For light treatments, a plate of dark-grown seedlings was exposed
to light ($150\,\mu\mathrm{E}$ m-2 s-1 PAR) for $4\,\mathrm{hours}$
while a control plate was kept in the dark, both plates being maintained
in a vertical orientation. To study the effect of Latrunculin B, $5$
control and $5$ treated cells were imaged. For the analysis of the
effect of light on the cytoskeleton, $35$ control and $26$ treated
cells were imaged. To fix the seedlings and to avoid mechanical damage,
they were mounted between a cover glass and a $1\,\mathrm{mm}$ thick
$1\mathrm{\%}$ agar pad affixed on a circular cover slip. A detailed
description of the microscopy setup is given in ref.~\cite{Sampathkumar2011}.
Typical exposure times were $400\,\mathrm{ms}$ for GFP and $300\,\mathrm{ms}$
for mCherry with a time interval of $2\,\mathrm{s}$ between subsequent
actin and microtubule images, respectively. The cells were recorded
for at least $4\,\mathrm{min}$. Only seedlings expressing both fluorescent
markers were used for further analyses.

\section{\label{sec:app_2}Network reconstruction procedure, different grid
topologies and null models}

\noindent \begin{flushleft}
\begin{figure*}
\begin{centering}
\includegraphics[width=1\textwidth]{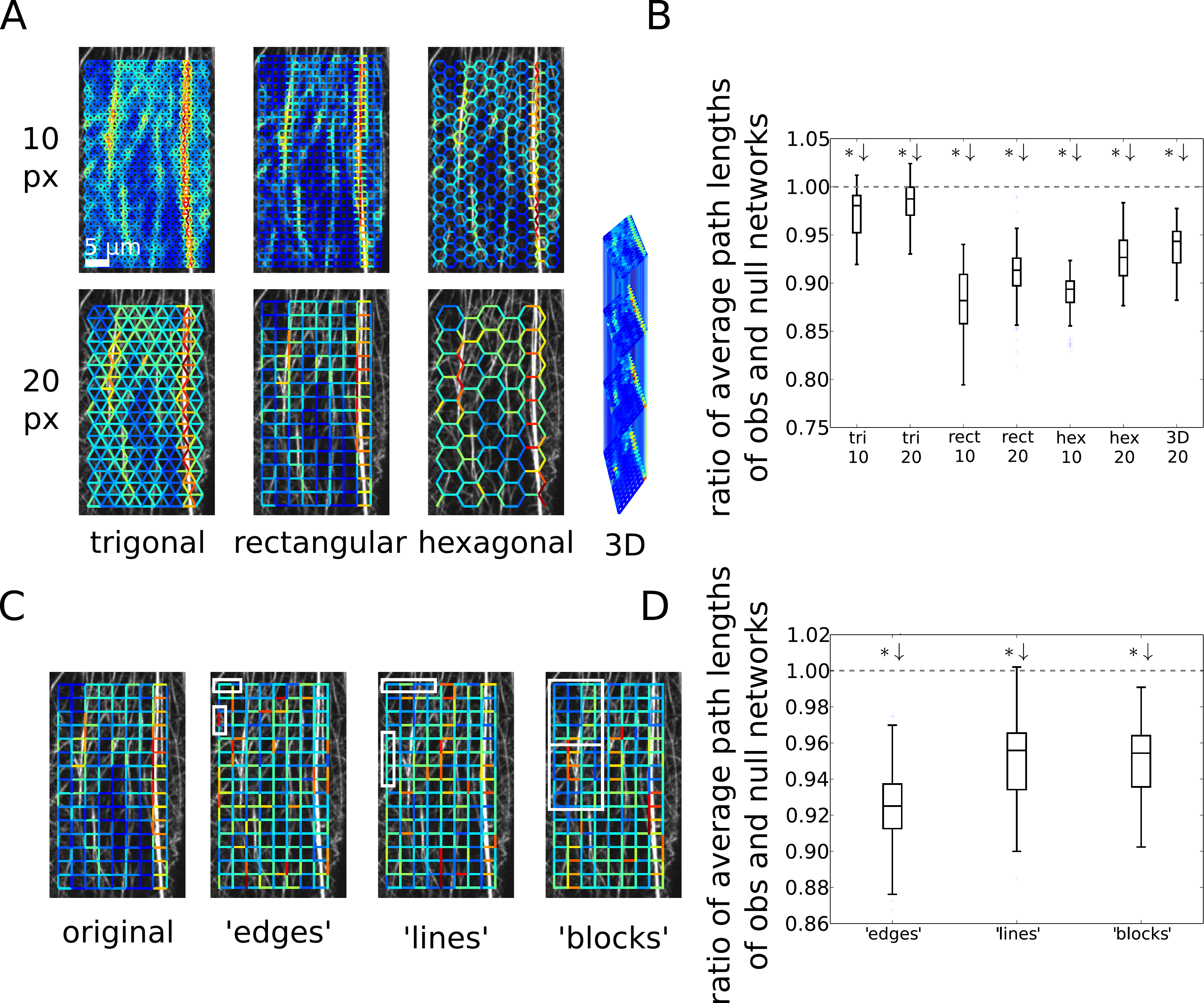}
\par\end{centering}

\protect\caption{\textbf{\label{fig:extensions}Extensions of the network reconstruction
framework and different null models.} \textbf{(A)} Reconstructions
of the actin cytoskeleton of the same cell using different sizes (grid
spacings of $10$ and $20\,\mathrm{pixels}$) of triangular, rectangular,
and hexagonal grids, and a three-dimensional grid with $20\,\mathrm{pixel}$-spacings.
\textbf{(B)} All studied grid types exhibit smaller average path lengths
than their respective null model networks. \textbf{(C)} Different
null models for the reconstructed network were obtained by shuffling
all edge weights, by shuffling connected vertical and horizontal lines,
and by rearranging blocks of varying size and shuffling the remaining
edge weights, respectively. White rectangles exemplify sections of
the cytoskeletal network that were shuffled. \textbf{(D)} All proposed
null models yield values below one for ratio of the average path lengths
of observed and null model networks, suggesting a non-random and efficient
organization of the cytoskeletal network.}
\end{figure*}

\par\end{flushleft}

We now explain our method for reconstructing a (edge-weighted, undirected)
network from the confocal image series of actin filaments or microtubules
and present various extensions and null models. We preprocessed the
recorded image series using Fiji \cite{Schindelin2012}: First, the
drift of the seedlings under the microscope was corrected using a
stack registration algorithm \cite{Arganda-Carreras2006}. Here, the
microtubule images were registered first because they are less dynamic
and easier for the program to align. The more dynamic actin filaments
were then subjected imagewise to the same transformations. Second,
the image series were rotated so that the shoot apical direction of
the cell pointed upwards. The region of interest, i.e., the interior
of the cell, was cropped manually and identically for the corresponding
actin and microtubule frames. Third, the background noise was reduced
by applying a rolling ball background subtraction \cite{Sternberg1983},
with a ball radius of $50\,\mathrm{pixels}$. The noisy background
signal arises largely from fluorescent monomeric actin/tubulin in
the cytosol which is not incorporated in filaments and, hence, was
filtered out in the present study. Finally, photobleaching was compensated
by rescaling all images' mean intensities to one.

From these preprocessed images we obtained the cytoskeletal components
as complex networks through a two-step procedure as described in the
main text (cf\emph{.}~section ``Reconstruction of complex networks
from cytoskeletal images''). These steps and all statistical analyses
were performed using Python \cite{VanRossum2013}. We chose an equidistant
rectangular grid with a spacing of $10\,\mathrm{pixels}$ and the
standard deviation of the Gaussian convolutions kernels was $4\,\mathrm{pixels}$
in x- and y-direction, unless stated otherwise. An extension of our
framework is the three-dimensional network that may be easily reconstructed
from three-dimensional confocal microscope image series (Fig.~\ref{fig:network}G).
First, a network was reconstructed for each z-slice as for the two-dimensional
case. Then, for a rectangular grid, the networks of neighboring z-slices
were connected by creating a link between nodes with the same x-y-coordinates.
The weights of these links were computed by creating Gaussian convolution
kernels (pointsymmetric, with the same width as for the edges, i.e.,
$4\,\mathrm{pixels}$) for its nodes, multiplying them with the two
respective z-slices and averaging over the sum of the resultant images.
The three-dimensional network reconstruction captures cytoskeletal
filaments which leave the cortical plane and may be analyzed using
the same network-based methods as for the two-dimensional networks.

To ensure that our findings (e.g., on the transport efficiency of
the cytoskeletal network architecture that displays short APLs; see
section \textquotedblleft Accessibility and robustness of cytoskeletal
networks\textquotedblright ), are valid not only for rectangular grids,
we tested other grid types. We reconstructed the cytoskeletal actin
network of the same cell based on two-dimensional rectangular, triangular,
and hexagonal grids with spacings of $10$ and $20\,\mathrm{pixels}$
and a three-dimensional grid with a uniform spacing of $20\,\mathrm{pixels}$
(Fig.~\ref{fig:extensions}A). For these networks, we compared the
APLs to an ensemble of null model networks obtained by edge-shuffling.
In all chosen grid types, the APL of the observed network is significantly
shorter than expected from the respective null model (Fig.~\ref{fig:extensions}B;
one-sample two-sided $t$-test: all $p\mathrm{-values}<0.05$). Hence,
the short APLs of the cytoskeleton are a non-random and biologically
relevant feature which does not arise as an artifact of the imposed
grid type. As network properties are often dependent on each other,
the findings from the comparative analysis suggest that a significant
change of other network properties compared to their null model values
is largely independent of the underlying grid type, as long as the
grid is not too dense, covers the cell too inhomogeneously (e.g.,
random geometric graphs), or has non-local, long-range links (e.g.,
scale-free graphs).

Judging the biological relevance of a network property's value requires
a meaningful comparison since its value depends on the normalization
of the image and is therefore arbitrary. The simplest reference is
given by the values of the respective network property which are obtained
for null model networks with shuffled edge weights (Fig.~\ref{fig:extensions}C,
``edges''). Such networks preserve both the node positions and the
distribution of edge weights and thus the total amount of cytoskeletal
components in the cell. By comparing the value of a given network
property of a reconstructed cytoskeletal network against those of
an ensemble of edge-shuffled null model networks, we were able to
assess whether a random distribution of cytoskeletal material in the
cell results in the same cytoskeletal properties as realized in the
observed cell, see section ``Accessibility and robustness of cytoskeletal
networks\textquotedblright . For example, for the APL, the ratio of
observed and null model values falls significantly below one (Fig.~\ref{fig:extensions}D,
``edges''; one-sample two-sided $t$-test: $p\mathrm{-value}<0.05$).

We also investigated two alternative null models to strengthen the
assessment of the biological relevance of different network properties.
Like the first null model, these, too, preserve the positions of the
nodes of the network and the distribution of edge weights in the network.
In addition, they leave more of the local cytoskeletal structure intact.
While in the first null model, edge weights were shuffled irrespective
of the edges to which they were assigned, we now cut all edges forming
connected horizontal or vertical lines into several equal sections,
respectively, which were in turn shuffled (Fig.~\ref{fig:extensions}C,
``lines''; here, horizontal lines are divided into three and vertical
lines in four sections). This method better preserves potential strongly
weighted paths and hence the filamentous structures of the cytoskeleton.
As for the ``edge\textquotedblright{} null model, this ``line\textquotedblright{}
null model exhibits longer APLs than the observed network (Fig.~\ref{fig:extensions}D,
``lines''; one-sample two-sided $t$-test: $p\mathrm{-value}<0.05$).
Clearly, the focus on horizontal or vertical lines imposes a restriction
to the orientation of potential filaments. To circumvent this limitation,
we analyzed a third null model in which connected, non-overlapping
blocks of nodes were chosen. The subgraphs formed by these blocks
were shuffled as well as the remaining edge weights that were not
part of any subgraph (Fig.~\ref{fig:extensions}C, ``blocks'';
here, the network is composed into three times four blocks). This
\textquotedblleft block\textquotedblright{} null model also exhibits
longer APLs than its biological counterpart (Fig.~\ref{fig:extensions}D,
``blocks''; one-sample two-sided $t$-test: $p\mathrm{-value}<0.05$).
More sophisticated null models may be proposed. However, the investigation
of three different null models that capture the amount of cytoskeletal
components in the cell and their filamentous structure provided consistent
results on the non-randomness of various cytoskeletal network properties.
Using the simple null model was, therefore, considered reliable for
assessing the biological relevance of the studied network properties
(see section \textquotedblleft Accessibility and robustness of cytoskeletal
networks\textquotedblright ).

\section{\label{sec:app_3}Network properties used for quantifying the cytoskeletal
organization}

\begin{figure*}
\begin{centering}
\includegraphics[width=1\textwidth]{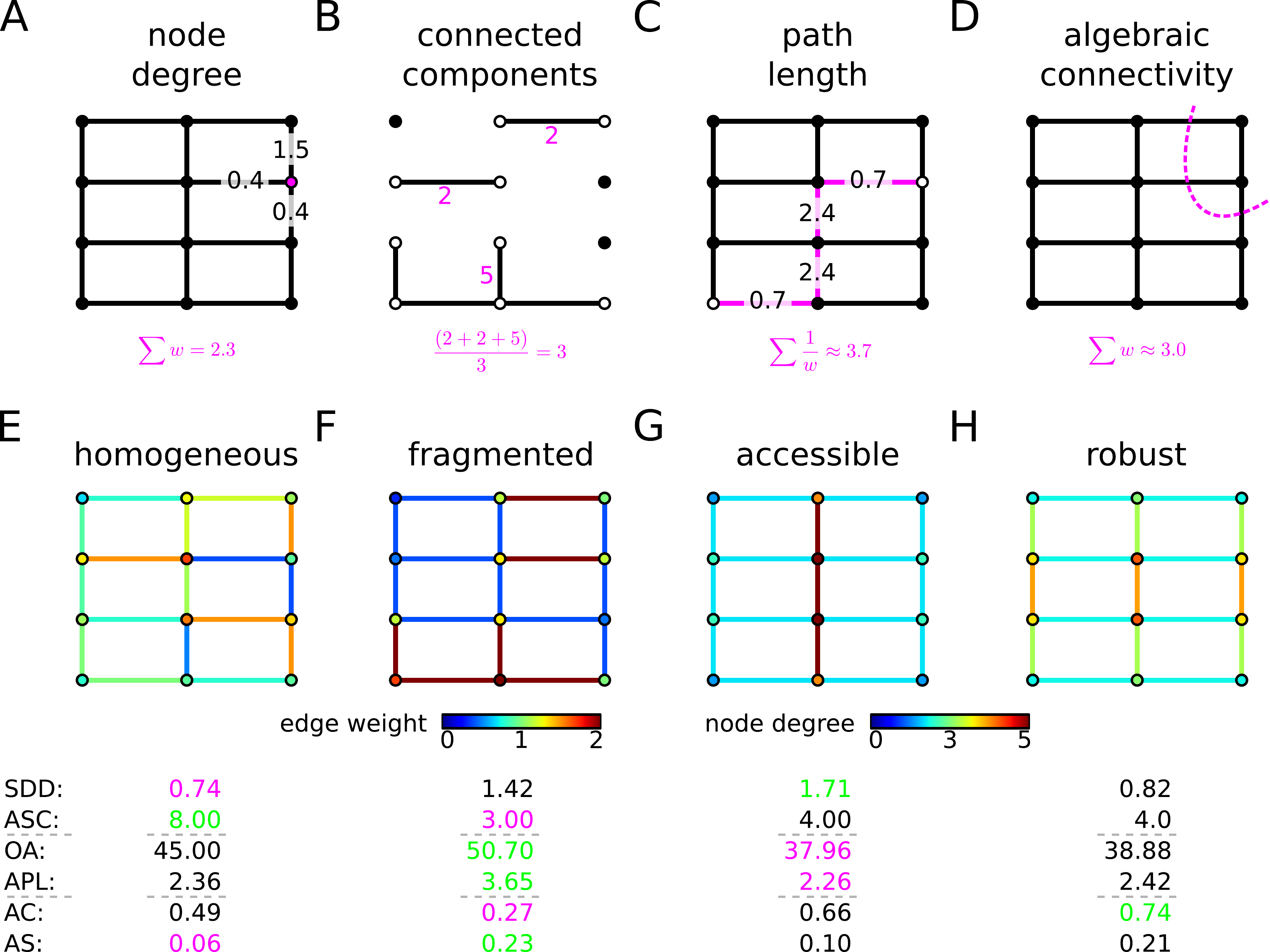}
\par\end{centering}

\protect\caption{\label{fig:schemes}\textbf{ Explanations of different network properties
and exemplary networks with different structural and functional characteristics.
}Panels (A)-(D)\textbf{ }display schematic illustrations of different
network properties. Panels (E)-(H) show typical paradigmatic networks
with color-coded edge weights and node degrees. \textbf{(A)} The degree
of a node is given by the sum of its edge weights, cf.~color-coding
in panel (E). \textbf{(B)} When removing edges with weights below
the 50th percentile from the network in panel (F), the network decomposes
into trivial components with just a single node (black circles) and
several bigger, non-trivial components (white circles). \textbf{(C)}
The shortest path length between two nodes is given by a sequence
of edges whose sum of inverse weights is minimal, cf.~network in
panel (G). \textbf{(D)} The algebraic connectivity relates to the
minimum sum of the weights of edges that need to be removed to disconnect
the network, cf.~network in panel (H). \textbf{(E)} The network exhibits
a small standard deviation of the degree distribution (SDD), has a
large average size of the connected components (ASC) after thresholding,
and a small assortativity (AS). \textbf{(F)} The network has a small
ASC, its overall angle (OA) indicates a horizontal orientation ($\mathrm{OA}>45^{\circ}$),
it displays a large average path length (APL), a small algebraic connectivity
(AC), and a large assortativity. \textbf{(G)} The network has a high
SDD, displays a vertical OA ($\mathrm{OA}<45^{\circ}$), and a small
APL. \textbf{(H)} The network was obtained by maximizing the AC for
a fixed sum of weights (cf\emph{.}~discussion of Fig.~\ref{fig:comparison})
and, accordingly, displays a high AC.}
\end{figure*}

In the manuscript, we represent the cytoskeleton as a weighted, undirected
network and quantify its structure via a number of seminal network
properties. Here, we explain the employed properties in more detail
and provide careful interpretations of how they relate to the structural
(and potentially: functional) features of the cytoskeleton. In the
following, we consider a weighted, undirected network $G=\left(\mathcal{N},\mathcal{E}\right)$
with a set $\mathcal{N}$ of $N=\left|\mathcal{N}\right|$ nodes,
a set $\mathcal{E}$ of $E=\left|\mathcal{E}\right|$ undirected edges
$e=\left(n,m\right)\equiv\left(m,n\right)\in\mathcal{E}$ and $m,n\in\mathcal{N}$
with weights $w_{e}$.
\begin{lyxlist}{00.00.0000}
\item [{\textbf{Degree~distribution:}}] The degree $d_{n}$ of a node
$n\in\mathcal{N}$ is given by the sum of its edge weights $w_{e}$,
i.e.\emph{,} 
\begin{eqnarray}
d_{n} & = & \sum_{\begin{array}{c}
e\in\mathcal{E}\\
n\in e
\end{array}}w_{e}\label{eq:prop_degree}
\end{eqnarray}
(cf\emph{.}~Fig.~\ref{fig:schemes}A for the degree of a node of
the network in panel E; color-coded node degrees in Figs.~\ref{fig:schemes}E-H).
Since the edge weights reflect the intensity of cytoskeletal structures
close to the respective edges, the node degrees reflect the cytoskeletal
intensities in the vicinity of the respective nodes. Therefore, the
standard deviation of the degree distribution (SDD) captures the spatial
heterogeneity of the distribution of intensities in the underlying
cytoskeleton images, 
\begin{eqnarray}
\mathrm{SDD} & = & \left(\left(N^{-1}\sum_{n=1}^{N}d_{n}^{2}\right)-\left(N^{-1}\sum_{n=1}^{N}d_{n}\right)^{2}\right)^{-1/2}.\label{eq:prop_sdd}
\end{eqnarray}
In particular, the SDD does not measure the heterogeneity in filament
thicknesses or numbers but a combination thereof. Comparing Figs.~\ref{fig:schemes}E
and G, we find that the SDD of the former network is smaller, indicating
a more homogeneous spatial distribution of the cytoskeleton in agreement
with the visual impression.
\item [{\textbf{Connected~components:}}] By construction, all edges in
the reconstructed networks are strictly positive (because the Gaussian
convolution kernels are strictly positive and have infinite support
and the image intensity is greater than zero somewhere in the image)
and, hence, all their nodes are connected. However, when removing
edges, e.g., with small weights by arguing that they do not permit
transport of cargo, the network may disconnect and split into several
connected components (cf\emph{.}~Fig.~\ref{fig:schemes}B for the
thresholded version of the network in panel F). The components are
called trivial if they consist of a single node only, and non-trivial
otherwise. For simplicity, we use the 50th percentile when thresholding
the edge weights throughout the manuscript but our findings remain
qualitatively unchanged when different, reasonable thresholds are
chosen. The average number of nodes per non-trivial connected component
(ASC) is a measure for the fragmentation of the network. 
\begin{eqnarray}
\mathrm{ASC} & = & C^{-1}\sum_{c=1}^{C}N_{c},\label{eq:prop_asc}
\end{eqnarray}
where $\mathcal{C}$ is the set of $C=\left|\mathcal{C}\right|$ non-trivial
components and $N_{c}$ is the number of nodes in component $c\in\mathcal{C}$.
Comparing Figs.~\ref{fig:schemes}E and F, we find that the ASC is
smaller in the latter which clearly exhibits several small, densely
connected fragments separated by weak connections.
\item [{\textbf{Overall~angle:}}] As our network representation does not
resolve individual filaments, we can not evaluate their orientations
individually. However, our approach allows to infer an overall angle
(OA) for the orientation of the cytoskeletal structures as a whole.
The OA is given by Eq.~\eqref{eq:angle_alpha} and its derivation
is explained in ESM4. Two networks with different OA are shown in
Figs.~\ref{fig:schemes}F and G with overall horizontal ($\mathrm{OA}>45^{\circ}$)
and vertical ($\mathrm{OA}<45^{\circ}$) orientations, respectively,
as confirmed visually.
\item [{\textbf{Average~path~length:}}] A path $\mathcal{P}$ between
two nodes is a sequence of edges connecting the nodes. A shortest
path is a path that minimizes its sum of edge lengths (cf\emph{.}~Fig.~\ref{fig:schemes}C
for a shortest path in the network in panel G). Here, for simplicity,
we take the length of an edge to be the inverse of its weight. This
choice takes into account that parts of the cytoskeleton that yield
strong edge weights potentially allow faster/more transport as reflected
by small edge lengths. 
\begin{eqnarray}
\mathrm{APL} & = & 2^{-1}N^{-1}\left(N-1\right)^{-1}\sum_{n=1}^{N}\sum_{\begin{array}{c}
m=1\\
m>n
\end{array}}^{N}\min_{\mathcal{P}\in\mathcal{P}_{n,m}}\sum_{e\in\mathcal{P}}w_{e}^{-1},\label{eq:prop_apl}
\end{eqnarray}
where $\mathcal{P}_{n,m}$ is the set of all paths from node $n$
to $m$ and $w_{e}^{-1}$ is the length of the edge $e\in P\in P_{n,m}$.
As explained in the discussion of Fig.~\ref{fig:efficiency} in the
main text, highway-like structures may yield small average path lengths
(APL) as they act as short cuts between distant parts of the network.
Such a highway-like structure is given by the network in Fig.~\ref{fig:schemes}G
which, accordingly, displays a smaller APL than, e.g., the easily-fragmented
network in panel F.
\item [{\textbf{Algebraic~connectivity:}}] The algebraic connectivity
is the second smallest eigenvalue 
\begin{eqnarray}
\mathrm{AC} & \equiv & \lambda_{2}\label{eq:prop_ac}
\end{eqnarray}
of the graph Laplacian $L$,
\begin{eqnarray}
L_{n,m} & = & \begin{cases}
d_{n} & \mathrm{if\,}n=m\\
-w_{\left(n,m\right)} & \mathrm{if\,}\left(m,n\right)\in\mathcal{E}\\
0 & \mathrm{otherwise},
\end{cases},\label{eq:prop_laplacian}
\end{eqnarray}
with $n,m\in\mathcal{N}$. By construction of $L$, its smallest eigenvalue
$\lambda_{1}=0$ and the number of zero eigenvalues provides the number
of connected components in the graph (cf.~e.g.~\cite{Newman2009a}).
As our reconstructed networks are always connected (see the discussion
of the connected components above) they yield $\lambda_{2}\equiv\mathrm{AC}>0$.
The magnitude of the $\mathrm{AC}$ is commonly interpreted as a measure
for how well-knit the network is, which is related to the minimum
sum of the weights of edges that need to be removed in order to disconnect
the network (cf\emph{.}~Fig.~\ref{fig:schemes}D). We solved a semi-definite
optimization problem described in the discussion of Fig.~\ref{fig:comparison}
to construct a network with a fixed sum of edge weights that maximizes
the $\mathrm{AC}$. This network is shown in Fig.~\ref{fig:schemes}H
and its $\mathrm{AC}$ is larger than, e.g., that of the easily-fragmented
network in panel F. \\
We note that small $\mathrm{APL}$ and large $\mbox{AC}$ favor different
types of networks (cf\emph{.}~Figs.~\ref{fig:schemes}G and H).
This may be explained as follows: In the computation of the $\mathrm{APL}$
for each shortest path only one edge may be used at a time. In contrast,
the $\mathrm{AC}$ is related to cuts (i.e.\emph{,} the removal of
sets of edges) that disconnect the network and hence affect multiple
edges. Thus, the $\mathrm{APL}$ and the $\mbox{AC}$ are independent
network properties that provide insights into different potential,
transport-related functions of the cytoskeleton.
\item [{\textbf{Assortativity:}}] The assortativity denotes the correlation
of the degrees of neighboring nodes
\begin{eqnarray}
\mathrm{AS} & = & \frac{1}{2E}\sum_{n=1}^{N}\sum_{m=1}^{N}\left(w_{\left(n,m\right)}-\frac{d_{n}d_{m}}{2E}\right)d_{n}d_{m}.\label{eq:prop_as}
\end{eqnarray}
Similar to the SDD (see above) the AS captures the spatial heterogeneity
of the cytoskeletal components but contains additional information
about its spatial distribution: The AS is high if nodes with high
(low) degrees are also connected to nodes with similar degrees, hence
detecting regions of spatially clustered cytoskeletal structures.
For instance, Fig.~\ref{fig:schemes}E displays a network with low
AS because the are no regions of nodes of high or low degree clustered
together, while the network in panel F shows high AS values that reflect
regions of high and low node degrees, respectively.
\end{lyxlist}

\section{\label{sec:app_4}Method for determining angles and filament orientations
from network structure}

\noindent \begin{flushleft}
\begin{figure*}
\begin{centering}
\includegraphics[width=1\textwidth]{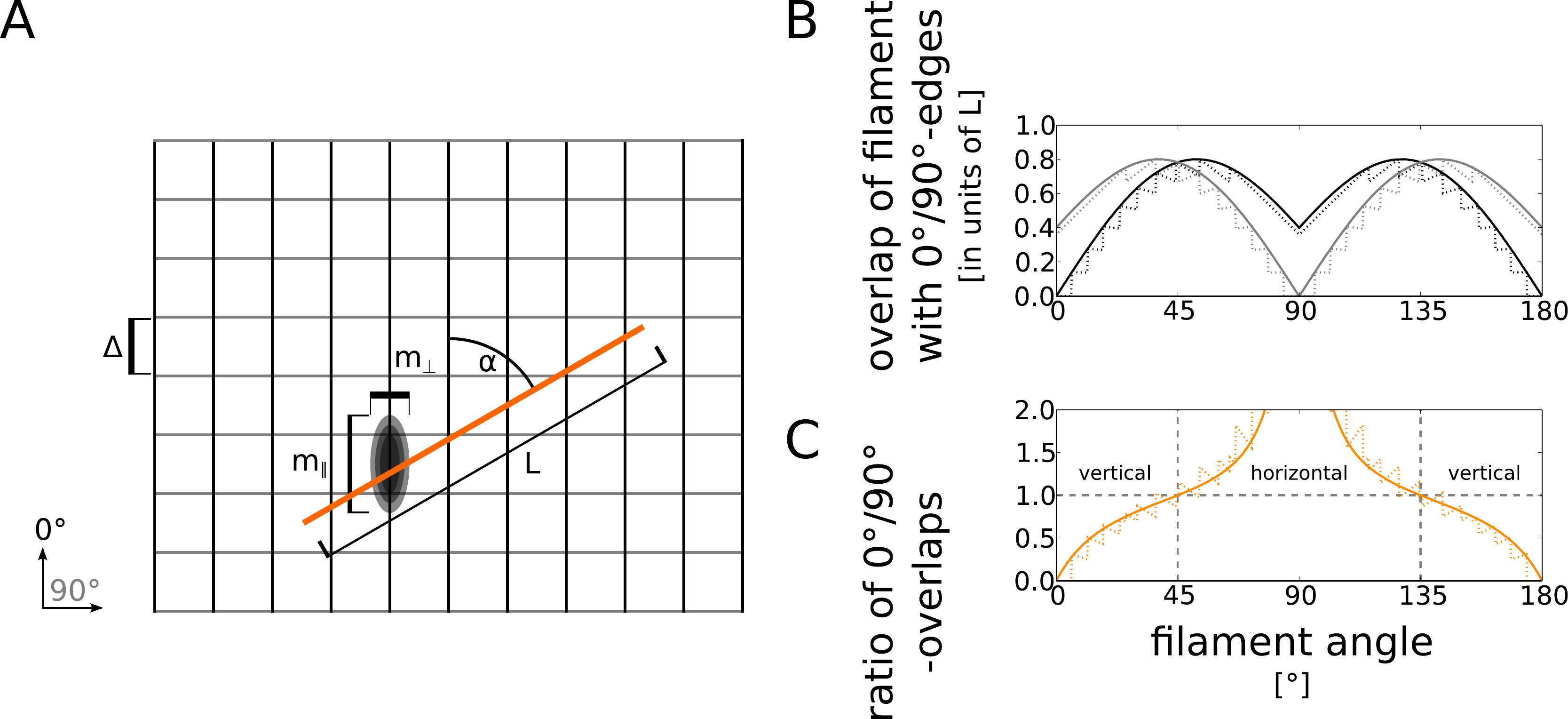}
\par\end{centering}

\protect\caption{\textbf{\label{fig:angles}Inferring orientation of filament from
network structure.} \textbf{(A)} Schematic filament of length $L$
at an angle $\alpha$, placed on a rectangular grid with spacing $\Delta$,
vertical ($0^{\circ}$, black), and horizontal edges ($90^{\circ}$,
gray). The shape of the edges\textquoteright{} convolution kernels
determines the contribution of the filament to each edge, with the
two extremes $m_{\parallel}$ and $m_{\perp}$ for parallel and perpendicular
orientation of the filament with respect to the edge. \textbf{(B)}
Contribution of filaments ($L=1000$, solid lines; $L=10$, dotted
lines) to edges of different orientation ($0^{\circ}$, black; $90^{\circ}$,
gray) for varying angles $\alpha$. By including that filaments cross
only integer numbers of edges, we obtain curves that are strongly
discontinuous for short filaments and become less discontinuous for
longer filaments. \textbf{(C)} The ratio of filament overlaps r with
$0^{\circ}$- and $90^{\circ}$-edges determines the filament angle
($L=1000$, solid line; $L=10$, dotted line), with $r<1$ and $r>1$
corresponding to a vertical and horizontal orientation, respectively.}
\end{figure*}

\par\end{flushleft}

We now present a method to evaluate the orientation of AFs and MTs
by exploiting their network structures. Starting from a given grid
(Fig.~\ref{fig:angles}A), we placed a stiff rod of length $L$ that
is rotated by an angle $\alpha$ and computed its contribution to
the weight of edges with an orientation of angle $\gamma$. Assuming
a regular grid in which edges with angle $\gamma$ are distributed
with uniform distances $\Delta$, we calculated the number $n_{\gamma}$
of crossed $\gamma$-edges,

\begin{eqnarray}
n_{\gamma} & = & \Delta^{-1}L\left|\sin\left(\alpha-\gamma\right)\right|.\label{eq:angle_number}
\end{eqnarray}
The overlap m of the rod and a $\gamma$-edge was computed via the
convolution kernel of that edge (see section ``Network reconstruction
procedure, different grid topologies and null models'') and was approximated
as

\begin{eqnarray}
m_{\gamma} & = & m_{\mathrm{\perp}}+\left(m_{\parallel}-m_{\perp}\right)\left|\cos\left(\alpha-\gamma\right)\right|,\label{eq:angle_overlap}
\end{eqnarray}
where $m_{\mathrm{\perp}}$ and $m_{\parallel}$ are the contributions
of the rod to the edge when they are perpendicular or parallel to
each other. The total contribution of the rod to all $\gamma$-edges
is (Eqs.~\eqref{eq:angle_number} and \eqref{eq:angle_overlap})

\begin{eqnarray}
w_{\gamma} & = & n_{\gamma}m_{\gamma}=\Delta^{-1}L\left|\sin\left(\alpha-\gamma\right)\right|\left[m_{\mathrm{\perp}}+\left(m_{\parallel}-m_{\perp}\right)\left|\cos\left(\alpha-\gamma\right)\right|\right].\label{eq:angle_weight}
\end{eqnarray}
Furthermore, we may include that only integer numbers of edges may
be crossed by a filament. Then, $L\sin$ in Eq.~\eqref{eq:angle_weight}
is replaced by $\left\lfloor L\sin\right\rfloor $ and the contribution
of a filament of finite length becomes discontinuous and approaches
the continuous curve for long filaments (Fig.~\ref{fig:angles}B).
For simplicity, we considered small grid spacings or long filaments
and work with Eq.~\eqref{eq:angle_weight} directly. Each cytoskeleton
fluorescence image contains many filaments and their lengths and orientations
can not be inferred uniquely from the distribution of edge weights.
Because $m_{\mathrm{\perp}}$ and $m_{\parallel}$ are determined
by the convolution kernels, the ratio of the weights of two edge types
with different orientations $\gamma$ and $\gamma^{'}$ yields an
equation for $\alpha$,

\begin{eqnarray}
\frac{w_{\gamma}}{w_{\gamma^{'}}} & = & \frac{\left|\sin\left(\alpha-\gamma\right)\right|\left[m_{\mathrm{\perp}}+\left(m_{\parallel}-m_{\perp}\right)\left|\cos\left(\alpha-\gamma\right)\right|\right]}{\left|\sin\left(\alpha-\gamma^{'}\right)\right|\left[m_{\mathrm{\perp}}+\left(m_{\parallel}-m_{\perp}\right)\left|\cos\left(\alpha-\gamma^{'}\right)\right|\right]}=:r,\label{eq:angle_ratio}
\end{eqnarray}
where $\alpha$ may be interpreted as the overall orientation of the
cytoskeletal filaments (Fig.~\ref{fig:angles}C). For $\gamma=0^{\circ}$
and $\gamma^{'}=90^{\circ}$, Eq.~\eqref{eq:angle_ratio} yields

\begin{eqnarray}
\alpha & = & \arctan\left(\frac{m_{\mathrm{\perp}}+r\left(m_{\parallel}-m_{\perp}\right)}{m_{\mathrm{\perp}}r+\left(m_{\parallel}-m_{\perp}\right)}\right).\label{eq:angle_alpha}
\end{eqnarray}
In our analysis, we refer to $\alpha\in\left[0^{\circ},45^{\circ}\right)$
and $\alpha\in\left(45^{\circ},90^{\circ}\right]$ as an overall vertical
and horizontal orientation, respectively. See section \textquotedblleft The
reconstructed networks capture biologically relevant features of the
actin and microtubule cytoskeletal components\textquotedblright{}
for results on the orientation of the cytoskeletal components under
different conditions.

\section{\label{sec:app_5}Reconstruction and analysis of the German autobahn
network}

\noindent \begin{flushleft}
\begin{figure}[H]
\begin{centering}
\includegraphics[width=1\textwidth]{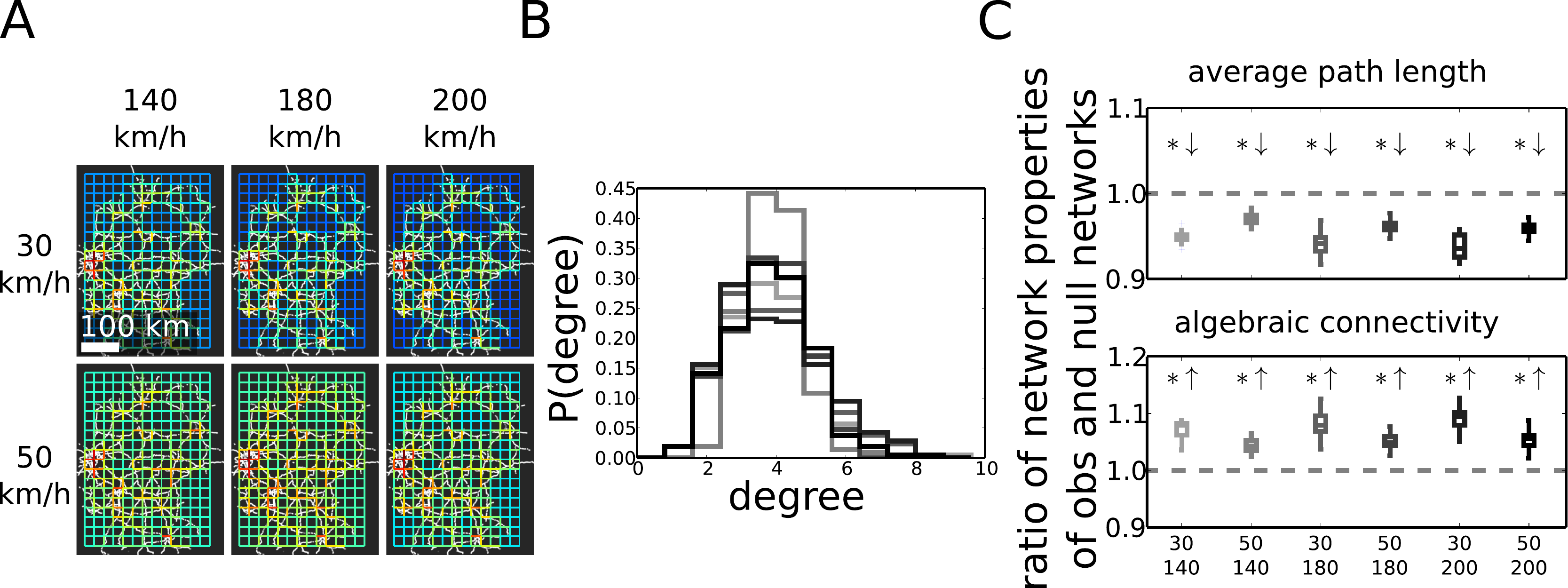}
\par\end{centering}

\protect\caption{\textbf{\label{fig:speedlimits}The efficiency of the autobahn network
is pertained for different choices of speed limits.} \textbf{(A)}
Reconstructed autobahn networks with different maximum high speed
limits ($140$, $180$, and $200\,\mathrm{km/h}$) and different off-highway
speed limits ($30$ and $50\,\mathrm{km/h}$). \textbf{(B)} The degree
distributions of the autobahn networks with the different speed limits
given in (A) (from light gray to black) are unimodal and centered
around their means (excess kurtosis $>0$). \textbf{(C)} The ratios
of average path lengths and algebraic connectivities of observed and
null model networks are well below and above one, respectively, for
all autobahn networks with speed limits described in (A).}
\end{figure}

\par\end{flushleft}

Here, we describe the data and the procedure used for reconstructing
the German autobahn as a network for a comparison with the plant cytoskeleton.
Further, we present two examples of networks with different structural
and transport-related properties.

An OpenStreetMap ($\textcircled{c}$ OpenStreetMap contributors; map
data available under the Open Database License (ODbL)) data set of
Germany was downloaded \linebreak (http://download.geofabrik.de/europe/germany.html;
$\textcircled{c}$ Geofabrik GmbH Karlsruhe), converted to~.o5m for
faster filtering (http://wiki.openstreetmap.org/wiki/Osmconvert),
and filtered for objects of type \textquotedblleft highway=motorway\textquotedblright{}
(http://wiki.openstreetmap.org/wiki/Osmfilter). The remaining motorways
were parsed in Python. Speed limits were take into account for better
analogy with the cytoskeleton that exhibits thinner and thicker bundles
that were argued in the main text to allow different net transportation
speeds. Because some sections of the autobahn were assigned no speed
limits (either because of lacking data or the absence of a speed limit)
and to incorporate transportation outside of the autobahn, we chose
different settings to ensure the robustness of our findings: Missing
autobahn speed limits were set to $140$, $180$, and $200\,\mathrm{km/h}$
and the speed limit in the rest of Germany was set to $30$ and $50\,\mathrm{km/h}$,
respectively (Fig.~\ref{fig:speedlimits}A).

The results for speed limits of $200\,\mathrm{km/h}$ and $50\,\mathrm{km/h}$
(Fig.~\ref{fig:comparison}; section ``The cytoskeleton and the
German autobahn exhibit similar network properties'') demonstrate
that the autobahn network displays a unimodal degree distribution
that peaks around its mean and that it exhibits significantly shorter
path length and a significantly higher AC than the null model networks.
The same holds true for all considered speed limits (Fig.~\ref{fig:speedlimits}B
and C; one-sample two-sided $t$-test: all $p\mathrm{-values}<0.05$).
Hence, the findings on the efficiency of the autobahn networks are
robust against moderate changes of the speed limits.
\end{document}